# Spin-polarized Current-driven Ferromagnetic Domain Wall Motion with a Skyrmion Building Block

O. Gorobets[1,2*], Yu. Gorobets[1,2], I. Tiukavkina[1], R. Gerasimenko[1]

[1]*National Technical University of Ukraine «Igor Sikorsky Kyiv Polytechnic Institute», 37 Peremohy Ave., 03056 Kyiv, Ukraine*

[2]*Institute of Magnetism of NAS and MES of Ukraine, 36b Acad. Vernadskoho Blvd., 03142 Kyiv, Ukraine*

*Correspondent author e-mail: gorobets.oksana@gmail.com, National Technical University of Ukraine «Igor Sikorsky Kyiv Polytechnic Institute», 37 Peremohy Ave., 03056 Kyiv, Ukraine

**Abstract**

The purpose of the research is the construction of the analytical model for description of spin-polarized current-driven ferromagnetic domain wall motion with a skyrmion building block. The dependence of velocity of ferromagnetic domain wall motion with a skyrmion building block is found as a function of driving torques and an external magnetic field strength.

1. **Introduction**

Recently, domain walls in ferromagnetic nanosized samples have been an urgent object of research as promising carriers of information bits for applications in magnetic memory devices [1]. Moreover, the domain wall in a ferromagnet can have the simplest defect-free structure, such as a "transverse" wall, or include vortices and other topological defects, such as a "vortex" wall. Among the wide variety of magnetic topological objects, the following are most distinguished as "building blocks" in the internal structure of domain walls: vortex, antivortex, bimeron, Bloch line, Bloch point.

In magnets, the Bloch lines divide the surface of a domain wall into two subdomains and significantly affect the properties of domain walls. Numerous studies have been devoted to the construction of magnetic memory devices based on Bloch lines [2,3]. To date, the Bloch lines are the most studied in ferromagnets with high uniaxial anisotropy [4]. The corresponding studies began much earlier in weakly anisotropic films [5], and modern achievements in this direction are described in [6]. The Bloch lines are observed regardless of the sign of the magnetic anisotropy constant in cubic ferromagnets [7,8]. A theoretical model of Bloch lines in weak ferromagnets was proposed in [9]. Local bends were observed in the Bloch lines moving at high speeds in yttrium orthoferrite, and they were associated with the movement of vortices along the domain wall [10].

The Bloch point is one of the examples of point topological defects in domain walls and was first proposed in [11,12]. The defining property of the Bloch point is that it represents the topological singularity of the magnetization field, and one can find all possible directions of the magnetization vector on the sphere of infinitely small radius centered at the Bloch point. Unlike other topological spin textures, such as magnetic skyrmions and vortices, the Bloch points have a unique feature – the local magnetization at the Bloch point completely disappears. This was experimentally confirmed in yttrium ferrite garnet crystals, micron thick garnet films, and magnetic cylindrical wire based on static measurements [13–15].

It was shown in [11] that the structure of the Bloch point is mainly determined by the exchange energy. Later in [12], the specific energy was calculated and it was shown that its value is topologically invariant. A family of magnetization textures with a local rotation angle γ (in the azimuthal direction) was considered in [12] and it was found that minimizing the magnetostatic energy selects a specific angle γ ≈ 112 ° [12]. In order to study the region near the singular point, the Landau magnetic energy [16] was included in [17] and the neglecting of magnetostatic energy was justified, and as a result, it was shown that the magnitude of the magnetization vector increases linearly with the radial distance from the center. The magnetization field of the Bloch point was calculated taking into account the exchange energy, Landau magnetic energy, and magnetostatic energy in [18]. The Bloch point in the domain wall of a ferromagnet is characterized by a topological (skyrmionic) charge q = ± 1 [19]. There are an infinite number of Bloch point configurations. However, there are three main possible configurations of the Bloch point, namely, a hedgehog configuration in which the magnetization distribution around the Bloch point is spherically symmetric, and the magnetization vector is directed away from the Bloch point (diverging Bloch point q = +1) or to the Bloch point (converging Bloch point q = −1), vortex or antivortex (q = +1 or −1) and spiral (q = + 1 or −1) configurations, which are obtained by 90 ° and 180 ° rotation of the magnetization of the similar configuration, respectively [20–22]. Direct observation of the stabilized structures of Bloch points with a skyrmion charge q = + 1, namely, hedgehog-like, vortex and spiral configurations, is reported in [19]. The in-plane and the out of plane magnetization components were observed using magnetic transmission soft X-ray microscopy MTXM [19] and the corresponding structures were determined based on numerical micromagnetic simulation [19].

"Vortex" or "topological" domain walls with integer topological charges, as well as integer or fractional winding numbers of volume vortices and edge defects are observed, for example, in ferromagnetic nanowires and nanorings [1]. The dynamics of a domain wall in a ferromagnet depends on the topological charge of the vortices in its structure. The movement of the domain wall leads to the creation, propagation, and annihilation of such defects.

The interest to the dynamics of magnetic vortices and Bloch points is also associated with the discovery of the fast magnetization reversal of the core of a magnetic vortex by alternating external influences (magnetic field [23] or spin current [24]). Numerical micromagnetic simulation of the magnetization reversal of the vortex core [25] showed that the annihilation mechanism of the vortex – antivortex pairs [26] requires mediation of the magnetization singularity: the "magnetic monopole" or, in other words, of the Bloch point [21].

At the same time, the vast majority of theoretical studies on the internal structure of domain walls in ferro- and antiferromagnets are based on numerical micromagnetic modeling. However, the results of the analysis of exact analytical solutions of the Landau-Lifshitz equations in ferro- and antiferromagnets have shown [27] that there can exist an infinite number of magnetic textures under the same boundary conditions for the magnetization vector (as well as the antiferromagnetism vector), which obviously represents problem for numerical micromagnetic simulation.

Therefore, in this work, we will obtain the exact dynamic solution of the Landau–Lifshitz–Gilbert–Slonczewski equation in a ferromagnet with uniaxial magnetic anisotropy, which describes the motion of a domain wall with a skyrmion in the internal structure under the influence of an external magnetic field and spin current.

## 2. Theory and calculation

Let us consider a ferromagnet with uniaxial magnetic anisotropy and magnetization $\vec{M}$, $|\vec{M}| = M_0$ where absolute value of magnetization is equal to $M_0 = const$. The expression for the magnetic energy of a ferromagnet and an equation of magnetization dynamics can be written though the angular variables that are introduces by the standard way:

$$M_x = M_0 \sin\theta \cos\varphi, \quad M_y = M_0 \sin\theta \sin\varphi, \quad M_z = M_0 \cos\theta, \tag{1}$$

where $\theta$ and $\varphi$ are the polar and azimuth angles for the magnetization, $M_x$, $M_y$, $M_z$ are the Cartesian coordinates of the magnetization vector.

The magnetic energy of a ferromagnet has the form

$$W = M_0^2 \int d\vec{r} \left\{ \frac{\alpha_{ex}}{2} \left[ \left(\frac{\partial \theta}{\partial x_i}\right)^2 + \sin^2 \theta \left(\frac{\partial \varphi}{\partial x_i}\right)^2 \right] + \frac{\beta}{2} \sin^2 \theta - \frac{H_0}{M_0} \cos \theta \right\}, \quad (2)$$

where $\alpha$ is the nonuniform exchange constant ($\alpha > 0$), $\beta$ is the constant of uniaxial magnetic anisotropy, $\vec{H}_0$ is an external magnetic field strength, integration in (2) is taken over the volume of a ferromagnet.

The Landau–Lifshitz–Gilbert–Slonczewski equation for a ferromagnet has the form

$$\frac{\partial \vec{M}}{\partial t} = -|g|\left[\vec{M} \times \vec{H}^{eff}\right] + \frac{\alpha_G}{M_0}\left[\vec{M} \times \frac{\partial \vec{M}}{\partial t}\right] + |g|\vec{T}, \quad (3)$$

where $\vec{T} = \vec{T}_\| + \vec{T}_\perp$ is the spin-transfer torque, i.e. the torque induced upon the magnetization by spin-polarized current flowing through the ferromagnet, $\vec{T}_\| = -\frac{a_J}{M_0}\left[\vec{M} \times \left[\vec{M} \times \vec{m}_p\right]\right]$, $\vec{T}_\perp = b_J\left[\vec{M} \times \vec{m}_p\right]$, $|g|$ is the gyromagnetic ratio, $\alpha_G$ is the damping factor, $\vec{H}^{eff}$ is the effective field, $\vec{H}^{eff} = -\frac{\delta W}{\delta \vec{M}}$, $a_J$ and $b_J$ are the driving torques, and $\vec{m}^p$ is the unit vector along the polarization of the current.

The Landau–Lifshitz–Gilbert–Slonczewski equation for a ferromagnet can be written though the angular variables

$$\begin{cases} \sin\theta\dfrac{\partial\varphi}{\partial t} = \dfrac{|g|}{M_0}\dfrac{\delta W}{\delta\theta} + |g|b_J m_z^p\left(-\sin\theta m_z^p + \sin\varphi\cos\theta m_y^p + \cos\varphi\cos\theta m_x^p\right) - \\ -|g|a_J\left(\sin\varphi m_x^p - \cos\varphi m_y^p\right) + \alpha_G\dfrac{\partial\theta}{\partial t} \\ -\sin\theta\dfrac{\partial\theta}{\partial t} = \dfrac{|g|}{M_0}\dfrac{\delta W}{\delta\varphi} + |g|b_J\sin\theta\left(\cos\varphi m_y^p - \sin\varphi m_x^p\right) + \\ -|g|a_J\left(\sin\theta\cos\theta\cos\varphi m_x^p + \sin\theta\cos\theta\sin\varphi m_y^p - \sin^2\theta m_z^p\right) + \alpha_G\sin^2\theta\dfrac{\partial\varphi}{\partial t} \end{cases} \quad (4)$$

The Landau–Lifshitz–Gilbert–Slonczewski equation can be simplified considering spin-torque polarization along OZ axis $m_x^p = m_y^p = 0,\ m_z^p = \pm 1$

$$\begin{cases} \sin\theta\dfrac{\partial\varphi}{\partial t} = \dfrac{|g|}{M_0}\dfrac{\delta W}{\delta\theta} - |g|b_J\sin\theta m_z^p + \alpha_G\dfrac{\partial\theta}{\partial t} \\ \sin\theta\dfrac{\partial\theta}{\partial t} = -\dfrac{|g|}{M_0}\dfrac{\delta W}{\delta\varphi} - |g|a_J\sin^2\theta m_z^p - \alpha_G\sin^2\theta\dfrac{\partial\varphi}{\partial t} \end{cases}. \quad (5)$$

It is possible to obtain the following equations for the magnetization dynamics substituting the energy of a ferromagnet with uniaxial magnetic anisotropy into the set of Landau–Lifshitz–Gilbert–Slonczewski equations

$$\begin{cases} \sin\theta\dfrac{\partial\varphi}{\partial t} = M_0|g|\left\{\alpha_{ex}\left[-\Delta\theta + \sin\theta\cos\theta(\nabla\varphi)^2\right] + \right. \\ \left. +\beta\sin\theta\cos\theta + \dfrac{H_0}{M_0}\sin\theta\right\} - |g|b_J\sin\theta m_z^p + \alpha_G\dfrac{\partial\theta}{\partial t} \\ \dfrac{\partial\theta}{\partial t} = M_0|g|\alpha_{ex}\left[\cos\theta\nabla\theta\nabla\varphi + \sin\theta\Delta\varphi\right] - |g|a_J\sin\theta m_z^p - \alpha_G\sin\theta\dfrac{\partial\varphi}{\partial t} \end{cases}. \quad (6)$$

### 3. Results and Discussion

The equations for the magnetization dynamics have the following exact dynamic solution

$$\begin{cases} \varphi = n\alpha + v_\varphi t + \alpha_0 \\ tg\dfrac{\theta}{2} = \exp\left(\dfrac{z - vt}{\delta} + n\ln\dfrac{r}{r_0}\right) \end{cases}, \qquad (7)$$

where

$$\begin{cases} v_\varphi = \dfrac{|g|H_0}{1+\alpha_G^2} - \dfrac{|g|m_z^p}{1+\alpha_G^2}(\alpha_G a_J + b_J) \\ v = \delta\left(\dfrac{\alpha_G}{(1+\alpha_G^2)}|g|H_0 + \dfrac{|g|m_z^p}{(1+\alpha_G^2)}(a_J - \alpha_G b_J)\right), \\ \delta = \sqrt{\dfrac{\alpha_{ex}}{\beta}} \end{cases} \qquad (8)$$

$n$ is an arbitrary integer number, $\alpha_0$ is an arbitrary initial phase, $0 \le \alpha_0 \le 2\pi$, $\delta$ is an arbitrary constant with the dimension of length.

This solution (7), (8) describes movement of a ferromagnetic domain wall of width $\delta$ with built in skyrmion with a constant velocity along OZ axis. The skyrmion as a domain wall building block can be of an arbitrary topological charge $n$ ($n$ is equal to both the skyrmion charge and the skyrmion winding number), of arbitrary size $\delta$ and of arbitrary initial helicity $\chi = \dfrac{\alpha_0}{n}$. The analytical model describes the temporary oscillations of skyrmion helicity from zero to $\chi = \dfrac{2\pi}{n}$. It means that during the period of such oscillations $T = \dfrac{2\pi}{v_\varphi}$ the skyrmion type is transforming from the Neel type at $\alpha_0 = 0$, $\alpha_0 = \pi$, $\alpha_0 = 2\pi$ to the intermediate type at arbitrary $\alpha_0 \ne 0$, $\alpha_0 \ne \pi$, $\alpha_0 \ne \pm\dfrac{\pi}{2}$, $\alpha_0 \ne 2\pi$, then to the Bloch type $\alpha_0 = \pm\dfrac{\pi}{2}$, and then again to the Neel type.

## 4. Conclusions

The exact dynamic solution (8), (9) of Landau–Lifshitz–Gilbert–Slonczewski equation in a ferromagnet with uniaxial magnetic anisotropy, obtained in the present paper, describe the spin-polarized current-driven ferromagnetic domain wall motion with a skyrmion building block. There is a linear dependence of velocity of ferromagnetic domain wall motion with a skyrmion building block $v$ as a function of driving torques and an external magnetic field strength according to the expression (8). The temporary oscillation of skyrmion type from Neel to Bloch one is predicted according to the formula (7) during the domain wall motion, the period of the oscillation is $T = \dfrac{2\pi}{v_\varphi}$.